\documentclass[english]{llncs}
\usepackage{amsmath,amssymb,babel}
\usepackage[final]{graphics,graphicx}
\usepackage{booktabs}
\usepackage{ifpdf}

\newcommand{\ceiling}[1]{\lceil {#1} \rceil}

\ifpdf
\else
\fi
\newcommand{\FPT}{\ensuremath{{\rm FPT}}}
\newcommand{\WSAT}{\ensuremath{W[{\rm SAT}]}}

\newcommand{\R}{\ensuremath{{\rm \bf R}}}
\newcommand{\N}{\ensuremath{{\rm \bf N}}}

\begin{document}

\title{Approximability and Parameterized Complexity of Minmax
  Values\thanks{Work supported by {\em Center for Algorithmic Game
      Theory}, funded by the Carlsberg Foundation.} \vspace*{-0.4cm}}
\author{Kristoffer Arnsfelt Hansen\thanks{Supported by a postdoc fellowship from the Carlsberg Foundation.}
\and Thomas Dueholm Hansen
\and Peter Bro Miltersen \and Troels Bjerre S{\o}rensen \vspace{-0.2cm}}
\institute{Department of Computer Science, University of Aarhus, Denmark\vspace{-0.3cm}
}

\maketitle

\begin{abstract}
\vspace*{-0.3cm}We consider approximating the minmax value of a multi-player
game in strategic form. Tightening recent bounds
by Borgs {\em et al.}, we observe that approximating the value with a
precision of ${\epsilon \log n}$ digits (for any constant $\epsilon>0$)
is {\bf NP}-hard, where $n$ is the size of the game. 
On the other hand, approximating the
value with a precision of $c \log \log n$ digits (for any constant $c
\geq 1$) can be done in quasi-polynomial time.
We consider the parameterized complexity of the problem, with
the parameter being the number of pure strategies $k$ of the player for which
the minmax value is computed. We show that if there are three
players, $k=2$ and there
are only two possible rational payoffs, the
minmax value is a rational number and can be computed {\em exactly}
in linear time. In the general case, we show that the value can be approximated
with any polynomial number of digits of accuracy in time $n^{O(k)}$. 
On the other hand, we show that
minmax value approximation is $W[1]$-hard and 
hence not likely to be fixed parameter
tractable. Concretely, we show that if
k-CLIQUE requires time $n^{\Omega(k)}$ then so does minmax value
computation.\vspace*{-0.3cm}
%
\end{abstract}

\section{Introduction}
\vspace*{-0.2cm}A {\em game $G$} in {\em strategic form} between $l$ players 
is given by a set of players $\{1,\dots,l\}$ and for each player $j$ a finite 
{\em strategy space} $S_j$ and a utility function 
$u_j: S_1 \times S_2 \times \cdots \times S_l \rightarrow \R$. In this paper, 
only the utility function for Player
1 is relevant. When the size of $S_j$ is $n_j$, we shall refer to
the game as an $n_1 \times n_2 \times \cdots \times n_l$ game.
The {\em minmax} (or {\em threat}) {\em value} of $G$ for 
Player 1 is given by
$\min_{\sigma_{-1}\in \Delta^{(l-1)}} \max_{a \in S_1} E[u_1(a, \sigma_{-1})]$
where $\Delta^{(l-1)}$ is the set of mixed, but uncorrelated, strategy
profiles for players $2, \ldots, l$. A profile $\sigma_{-1}$ achieving
the minimum in the expression is called an {\em optimal minmax profile} or
an {\em optimal threat}.
The {\em maxmin} (or {\em security}) {\em value} of $G$ for Player 1 is given by
$\max_{\sigma_1 \in \Delta} \min_{a_2, \ldots, a_l} E[u_1(\sigma_1,
a_2, \ldots, a_l)]$
where $\Delta$ is the set of mixed strategies
for Player 1.

The minmax value of a finite two-player game
 is a fundamental notion of game
theory. Its mathematical and computational properties are extremely 
well-studied and well-understood, being intimately tied to the
theory of linear programming. In particular, the duality theorem
of linear programs implies that the minmax value equals the 
maxmin value.
Also, the computation of the minmax value of a two-player
game in strategic form is essentially equivalent to solving linear
programs
and can therefore be done in polynomial time (although a strongly
polynomial time algorithm remains an open problem).

Minmax values of {\em multi}-player games are much less well-studied,
although these values are arguably also of fundamental interest to
game theory. Most importantly, the minmax value pays a pivotal role
in the statement and proof of the so-called {\em folk theorem}
that characterize the Nash equilibria of
infinitely repeated games. Additionally, the minmax value is the
equilibrium payoff of the so-called {\em team-maxmin equilibria}
studied by von Stengel and Koller \cite{GEB:KS97}.
For a multi-player game, the maxmin value may be strictly smaller
than the minmax value. Computation of the maxmin value easily
reduces to the two-player case and can therefore be done efficiently
using linear programming.
Rather surprisingly, computation of the minmax value of a
multi-player game in strategic form was not studied until very recently, where
Borgs {\em et al.} \cite{STOC:BCIKMP08} (motivated by computational aspects of the 
folk theorem) showed that approximating the minmax value of a
three-player game within a certain inverse polynomial additive error is 
${\bf NP}$-hard.
Our starting point is this important paper.

Given the fundamental nature of the notion of the minmax value, it is
important to understand when the {\bf NP}-hardness result can be
circumvented by considering special cases or asking for weaker
approximations. The purpose of this paper is to provide a 
number of results along these lines. 
First, we observe that the inapproximability result of Borgs
{\em et al.} can be tightened and matched with a positive result, 
using standard techniques:
\begin{theorem}\label{thm:lipton}
For any constant $\epsilon > 0$, approximating the minmax value of an 
$n \times n \times n$ game with 0-1 payoffs within additive error 
$1/n^{\epsilon}$ is {\bf NP}-hard. 
On the other hand there is an algorithm that, given a parameter $\epsilon > 0$ 
and a game in strategic form with $l$ players each having $n$ strategies and 
all payoffs being between $0$ and $1$, approximates the minmax value for 
Player 1 from above with additive error at most $\epsilon$ in time 
$n^{O(l(\log n)/\epsilon^2)}$.
\end{theorem}
This suggests the following important problem: {\em Can the minmax value of a
three-player
game with payoffs normalized to $[0,1]$ be approximated within 
a fixed constant (say 0.01 or even 0.99) in polynomial (rather than
quasi-polynomial) time?} We leave this problem open.
 
It is of interest to know when the minmax value can be computed
exactly. A prerequisite for this is that it is rational. 
For three-player games, we characterize when the minmax value for Player 1 can be an
irrational number, in terms of the number of strategies of Player 1
and the number of distinct (rational) payoffs. For the special case
where the value is guaranteed to be rational we  present an
optimal linear time algorithm for exactly computing the minmax value\footnote{
As the algorithms of Theorem \ref{thm:lipton} and Theorem
\ref{thm:linear} are very simple, we express their complexity in
the unit cost random access machine model. E.g., by ``linear time'' we
mean a linear number of atomic operations in the number of real
payoffs of the input. On the other hand, the algorithm of Theorem 
\ref{thm:support} use sophisticated algorithms from the literature as
subroutines and its complexity is better expressed in the Turing machine model,
and in terms of bit complexity.}.

\begin{theorem}\label{thm:linear}
  Consider $k \times n \times n$ three-player games with only $l$
  distinct rational payoffs. When either $k \geq 2$ and $l \geq 3$ or
  $k \geq 3$ and $l \geq 2$ there exists a game such that the minmax
  value for Player 1 is irrational. Otherwise, when $k=2$ and $l=2$
  the minmax value for Player 1 is a rational number and we can
  compute it exactly in time $O(n^2)$ (on a unit cost 
  random access machine model).
\end{theorem}

Thus having observed that the case of few strategies of Player 1
may be easier than the general case, we apply the
approach of {\em parameterized complexity} \cite{DowneyFellows99},
considering the number of strategies $k$ of Player 1
as the parameter. 
Combining a classical result of Shapley and Snow \cite{AMS:SS50} 
with Renegar's decision procedure for the first order theory of the reals
\cite{JSC:Renegar92a,JSC:Renegar92b,JSC:Renegar92c} gives rise to a
support enumeration algorithm for finding the minmax value and we show
\begin{theorem}\label{thm:support}
  Given a $k \times n \times \cdots \times n$ $l$-player
  game $G$ with rational payoffs and a
  rational number $\alpha$ so that $(G,\alpha)$ has combined bit
  complexity $L$, we can decide in time $L^{O(1)}
  k^{O(kl)} n^{k l}$ (on a Turing machine) if the minmax value of $G$
  for Player 1 is at most $\alpha$.  Using the terminology of fixed
  parameter complexity theory, considering $k$ the parameter,
  a modification of the algorithm shows that this problem
  is in $W[P]$, and for the case of 0-1 payoffs in $W[1]$.
\end{theorem}
In particular, if $l$ and $k$ are constants, the complexity is
polynomial, and we can approximate the minmax value with
any polynomial number of bits of accuracy in polynomial time by using the
decision procedure in a binary search.
As the exponent in the above complexity bound depends linearly on $k$
with impractical bounds for large $k$ as consequence,
we next ask if the problem of approximating
the minmax value for Player 1 in a three-player game is 
{\em fixed parameter tractable},
i.e., if an algorithm solving the problem in time $f(k) n^c$ exists,
where $f$ is any function and $c$ is a constant not depending on $k$.
We provide a reduction from $k$-\textsc{Clique} that gives negative
evidence. 

\begin{theorem}\label{thm:reduction}
Let $G = (V,E)$ be an undirected graph with $|V| = n$, then
$k$-\textsc{Clique} reduces to the problem of
approximating the threat value for Player 1 within
$1/(4k^2)$ in a three-player $2 k
\times k n \times k n$ game with payoffs $0$ and $1$.
\end{theorem}
Downey and Fellows \cite{TCS:DF95} proved that the \textsc{Clique}
problem is complete for the class $W[1]$, and hence it
immediately follows that the problem of approximating the
minmax value within $1/(4k^2)$ for Player 1 in a $2 k \times k n
\times k n$ game with $k$ being a parameter is hard for $W[1]$, even
when all payoffs are $0$ or $1$. Combining this with Theorem
\ref{thm:support}, we  in fact have that the 0-1 case is $W[1]$-complete.
Readers not well-versed in the theory of parameterized complexity may
find the following consequence of the reduction more appealing: The
minmax value of a $2k \times kn \times kn$ three-player game with 0-1
payoffs cannot be approximated in time $n^{o(k)}$, unless
$k$-\textsc{Clique} can be solved in time $n^{o(k)}$.  If 
$k$-\textsc{Clique} could be solved in time $n^{o(k)}$ then as proved
by Chen {\em et al.} \cite{IC:CCFHJKX05} it would follow that all
problems in the class {\bf SNP} (e.g. 3-SAT) could be solved in time
$2^{o(n)}$. Thus, under the assumption that all of {\bf SNP} cannot be
solved in time $2^{o(n)}$, the algorithm of Theorem \ref{thm:support} is
essentially optimal for the case of 0-1 payoffs, in the sense that 
its complexity is $n^{O(k)}$ and $n^{\Omega(k)}$ is a lower bound.

%
\vspace*{-0.2cm}

\section{Proofs}
\subsection{Proof of Theorem \ref{thm:lipton}}
We first prove the hardness claim.
Borgs {\em et al.} showed hardness of approximation with additive
error $3/n^2$. Now consider, for a positive integer $c \geq 2$ 
the following ``padding'' construction:
Given an $n \times n \times n$ game $G$ with strategy space $S_i$ of
Player $i$ and utility function $u_1$ for Player 1.  
Let $n' = n^c$ and define the 
$n' \times n' \times n'$ game $G'$ with strategy space
$S'_i = S_i \times \{1, \ldots, n^{c-1}\}$ and utility function for Player 1
being $u'_1((x,a_1), (y,a_2), (z,a_3)) = u_1(x,y,z)$. In words, $G'$ is simply
$G$ with each strategy copied $n^{c-1}$ times. Now, $G'$ and $G$ 
clearly have the same minmax value. Also, for a given $\epsilon > 0$, 
by picking $c$ to be a large enough constant, we can ensure that 
$1/(n')^{\epsilon} < 3/n^2$, so
approximating the minmax value of $G$ within $3/n^2$ reduces
to approximating the minmax value of $G'$ within $1/(n')^\epsilon$,
which concludes the proof of hardness
(we remark that this simple padding argument also yields a somewhat
simpler proof of Lemma 7.1 of Chen, Teng and Valiant \cite{SODA:CTV07}).

We now proceed with the positive approximation result. We only show
the result for the case of three players; the general case being very
similar.
For the proof, we will use the following Theorem by Lipton and Young
\cite[Theorem 2]{LipY}:
\begin{theorem}\label{LY}
For a two-player zero-sum $n \times n$ game with payoffs in $[0,1]$, there is
a simple strategy for each player that guarantees that player a payoff
within $\epsilon$ of the value of the game. Here, a simple strategy 
is one that mixes uniformly
on a multiset of $\ceiling{\ln n/(2\epsilon^2)}$ strategies.
\end{theorem}
Now consider a given $3$-player game $G$ and consider the optimal threat
strategy profile $(\sigma_2, \sigma_3)$ of Players $2$ and $3$ against 
Player $1$. Consider $\sigma_3$ as fixed and look at the resulting
two-player game $G'$ between Player 1 (maximizer) and Player 2
(minimizer). 
Clearly, this
game has value equal to the minmax value for Player $1$ in $G$.
Applying Theorem \ref{LY}, there is a simple strategy $\sigma'_2$
for Player 2 that guarantees this value within $\epsilon$.
Fix $\sigma'_2$ to this strategy and look at the resulting
two-player game $G''$ between Player 1 and Player 3. By construction
of
$\sigma'_2$, this
game has value at most $\epsilon$ larger than the value of $G'$.
Applying Theorem \ref{LY} again, there is a simple strategy $\sigma'_2$
for Player 2 that guarantees this value within $\epsilon$. Thus,
if Player 2 and Player 3 play the profile $(\sigma'_2, \sigma'_3)$,
in the original game, they are guaranteed the minmax value of $G$
plus at most $2 \epsilon$.

Now, given some $\epsilon'$, we let $\epsilon = \epsilon'/2$ and
approximate the threat value of
Player 1 within $\epsilon'$ by exhaustively searching through all
pairs of simple strategies for Player 2 and Player 3, compute the
maximum payoff for Player 1, and return the lowest such payoff. 
This completes the proof of the theorem.\\

It is natural to ask if one can get any non-trivial approximation
by considering strategies that mix uniformly over only a constant
size multiset, as this would lead to a polynomial time approximation
algorithm rather than a quasi-polynomial one. Unfortunately, the
answer is negative:
For given $n$ and $c$, let $m$ be maximal such that $\binom{m}{c}^2
\leq n$. Consider the $\binom{m}{c}^2 \times m \times m$ game $G$ 
defined as
follows. For every two subsets of actions of size $c$ for Player 2 and
Player 3 there is an action for Player 1 such that he receives payoff
1 if the other players chose an action from the two subsets. Otherwise
Player 1 receives payoff 0. If Player 2 and Player 3 choose their
actions uniformly at random, Player 1 can ensure payoff at most
$(\frac{c}{m})^2$, while on the other hand if the strategies of Player
2 and Player 3 has support size at most $c$, then Player 1 can ensure
payoff $1$. We can now, in a similar way as
in the proof of Theorem \ref{thm:lipton} above,
construct a padded version of
the game, obtaining a $n \times n \times n$ game $G'$ such that the
minmax value for Player 1 is at most $(\frac{c}{m})^2 \leq
\frac{(2e)^2}{n^{\frac{1}{c}}}$, but for every strategy profile for Player 2
and Player 3 of support size at most $c$, Player 1 can ensure payoff
1. Thus to approximate the minmax value within, say, $\frac{1}{2}$ we
must have $c \geq \frac{\ln n}{\ln(8e^2)}$.

\subsection{Proof of Theorem \ref{thm:linear}}
First, we give the claimed examples of games for which the minmax value
for Player 1 is irrational. We describe the games by a matrix for each
action of Player 1, where rows and columns correspond to the
actions of Player 2 and Player 3, respectively. That is, we let
$u_1(i,j,k)=A_i(j,k)$.

The first game is a $2 \times 2 \times 2$ game where there are $3$
distinct payoffs, given by the following matrices.
\[
A_1 = \left[\begin{array}{cc}
1 & 0\\
0 & 0
\end{array}\right]
A_2 = \left[\begin{array}{cc}
0 & 0\\
0 & 2
\end{array}\right]
\]
It is easy to see that the strategy for Player 2 and Player 3 yielding
the minmax value for Player 1 is to play the first action with
probability $2-\sqrt2$. This results in a minmax value of $6-4\sqrt2$
for Player 1.

The second game is a $3 \times 2 \times 2$ game where there are $2$
distinct payoffs, given by the following matrices.

\[
A_1 = \left[\begin{array}{cc}
1 & 0\\
0 & 0
\end{array}\right]
A_2 = \left[\begin{array}{cc}
0 & 0\\
1 & 1
\end{array}\right]
A_3 = \left[\begin{array}{cc}
0 & 1\\
0 & 1
\end{array}\right]
\]
It is easy to see that the strategy for Player 2 and Player 3 yielding
the minmax value for Player 1 is to play the first action with
probability $\frac{\sqrt 5 - 1}{2}$ (golden ratio conjugate).  This
results in a minmax value of $\frac{3-\sqrt 5}{2}$ for Player 1.

We now examine the special case where there are only two distinct payoffs, and 
Player 1 only has two possible actions.  For this case, we show that the 
threat point is always rational valued, and the threat strategies can be 
computed in linear time.  
We assume without loss of generality that
the two possible payoffs are 0 and 1.
The proof is a case analysis, where each case can 
be identified and solved in linear time, assuming  that none of the 
previous cases apply.  Case 1 and 2 are the trivial cases where either side 
has a pure optimal strategy. 

\noindent
{\bf Case 1: \boldmath $\exists i\forall j,k: u_1(i,j,k)=1$.} 
Player 1 has a ``safe'' action, $i$, such that no matter what Players 2 and 3 
does, Player 1 achieves value 1.  Any strategy profile for Players 2 
and 3 is an optimal threat, with minmax value 1.

\noindent
{\bf Case 2: \boldmath $\exists j,k\forall i: u_1(i,j,k)=0$.}
Players 2 and 3 have a pure strategy threat, such that no matter what Player 1 
does, the payoff is 0.  The strategy profile $(j,k)$ is an optimal threat, 
with minmax value 0.

\noindent
Case 1 and Case 2 can easily be identified and solved in linear time.
Notice that when we are not in case 2, we have that 
$\forall j,k:\exists i: u_1(i,j,k)=1$, and therefore 
$u_1(i,j,k)=0\Rightarrow u_1(i',j,k)=1$ where $i'\not=i$.  This means that Player 1 
has a maxmin (security) value of at least $\frac12$, which can be achieved by a 
uniform mix of the two strategies.  As the minmax (threat) value has to be 
greater than the maxmin value, any threat with minmax value $\frac12$ 
will be optimal.  This is exactly what can be achieved in the next two cases:

\noindent
{\bf Case 3: \boldmath $\exists j\forall i\exists k:u_1(i,j,k)=0$.}
Player 2 has a pure strategy, such that Player 3 can play matching pennies 
against Player 1.  Let $k$ and $k'$ be the strategies of Player 3 achieving 
payoff $0$ against $i$ and $i'$ respectively.  Then $(j,(\frac12k,\frac12k'))$ is 
an optimal threat, with minmax value of $\frac12$. 

\noindent
{\bf Case 4: \boldmath $\exists k\forall i\exists j:u_1(i,j,k)=0$.}
Player 3 has a pure strategy, such that Player 2 can play matching pennies 
against Player 1.  Let $j$ and $j'$ be the strategies of Player 3 achieving 
payoff $0$ against $i$ and $i'$ respectively.  Then $((\frac12j,\frac12j'),k)$  is 
an optimal threat, with minmax value of $\frac12$.

\noindent
Case 3 and Case 4 can again easily be identified and solved in linear time.

\noindent
{\bf Case 5: None of the above.}
The negation of case 1 implies $\forall i\exists j, k: u_1(i,j,k)=0$.
The negation of 
cases 2, 3 and 4 implies
$u_1(i,j,k)=0\Rightarrow\forall j',k':u_1(i',j',k)=u_1(i',j,k')=1$, where $i'\not= i$.  
That is, any strategy profile of Player 2 or 3 can achieve payoff 0 
against at most one of Player 1's strategies, and Players 2 and 3 must agree on which 
strategy of Player 1 to try to get payoff 0 against; if they disagree, the 
payoff is 1 no matter what Player 1 does.  The best they can hope for is 
therefore $\min_{p,q\in[0;1]}\max\{1-pq,1-(1-p)(1-q)\}$, which gives a lower 
bound on the minmax value of $\frac34$.  This value can be achieved in this 
case in the following way: let $u_1(i,j,k)=u_1(i',j',k')=0$.  Then 
$((\frac12j,\frac12j'),(\frac12k,\frac12k'))$  is an optimal threat, with 
minmax value of $\frac34$.

\subsection{Proof of Theorem \ref{thm:support}}
We prove the theorem for three players, the general case is similar.
Shapley and Snow \cite{AMS:SS50} showed that every $k \times n$
zero-sum game has a minmax mixed strategy for Player 2 of support at
most $k$, i.e., using at most $k$ strategies.  We claim that from this
it follows that in every $k \times n \times n$ game there are
strategies for Player 2 and Player 3 of support at most $k$ so that
the resulting strategy profile $\sigma_{-1}$ yields the minmax value
for Player 1 when Player 1 chooses a best response. Indeed, consider
the actual minmax strategy profile $\sigma_{-1}^* = (\sigma^*_2,
\sigma^*_3)$. If we consider $\sigma^*_3$ fixed and consider the
resulting two-player game between Player 1 and Player 2, it is clear
that $\sigma^*_2$ is a minmax strategy of this game and that Player 2
will still guarantee the minmax payoff by playing the minmax
strategy $\sigma^*_2$ of support $k$ which is guaranteed to exist by
Shapley and Snow's result. Similarly, we may replace $\sigma^*_3$ with
a strategy of support $k$ without changing the payoff resulting when
Player 1 plays a best response.

Our algorithm is a support enumeration algorithm which exhaustively
examines each possible support of size $k$ for Player 2 and Player
3. From the above observation it follows that the minmax value of the
game is the minimum of the minmax value of each of the resulting
$k \times k \times k$ subgames.  Therefore, we only have to explain
how to compare the minmax value of such a subgame to a given
$\alpha$, and we will be done.  For this, we appeal to classical
results on the first order theory of the reals.

The decision procedure for the first order theory of the reals due to
Renegar \cite{JSC:Renegar92a,JSC:Renegar92b,JSC:Renegar92c} can decide
a sentence with $\omega-1$ quantifier alternations, the $k$th block of
variables being of size $n_k$, containing $m$ atomic predicates and
involving only polynomials of degree at most $d$ with integer
coefficients of maximum bit length $L$ using $L(\log L) 2^{O(\log^*
  L)}(md)^{2^{O(\omega)}\prod_k n_k}$ bit operations\footnote{The
  bound stated here is the improvement of the bound stated by Renegar
  due to the recent breakthrough in integer multiplication due to
  F{\"u}rer \cite{STOC:Furer07}.} and $(md)^{O(\sum_k n_k)}$
evaluations of the Boolean formula of the atomic predicates.
We claim that from this it follows that
  given a $k \times k \times k$ game $G$ with rational payoffs and a
  rational number $\alpha$ so that $(G,\alpha)$ has combined bit
  complexity $L$, we can decide in time $L(\log L) 2^{O(\log^* L)}
  k^{O(k)}$ (on a Turing machine) if the minmax value of $G$ for
  Player 1 is at most $\alpha$.
  We can assume that the payoffs and $\alpha$ are integers at the
  expense of increasing the bitlength of every number to at most the
  combined bitlength of the original problem.
  Define the following polynomials in $2k$ variables.
\[
\begin{split}
&p_l(x_1,\dots,x_{2k}) = \sum_{i=1}^k \sum_{j=k+1}^{2k} u_1(l,i,j) x_ix_j \quad,\quad
r_i(x_1,\dots,x_{2k}) = x_i\\
&q_1(x_1,\dots,x_{2k}) = \sum_{i=1}^k x_i \quad,\quad
q_2(x_1,\dots,x_{2k}) = \sum_{i=k+1}^{2k} x_i
\end{split}
\]
The sentence we must decide is then
\[
(\exists x \in \R^{2k}) [p_1(x) < \alpha \wedge \dots \wedge
  p_k(x) < \alpha \wedge q_1(x)=1 \wedge q_2(x)=1 \]
\[
\wedge r_1(x) \geq 0 \wedge \dots \wedge r_{2k}(x)\geq 0 ].
\]
For this sentence we have $\omega=1$, $m=3k+2$, $d=2$ and $n_1=2k$,
and the sentence can thus be decided in the claimed running time using
Renegar's procedure.
For the support enumeration algorithm this decision procedure must be
invoked for $\binom{n}{k}^2$ different $k \times k \times k$ subgames,
and the claimed time bound of the statement of the theorem follows.

Next, we show how to use this algorithm to show $W[P]$ and $W[1]$
membership of the two versions of the problem. We use the framework
of aftp-programs of Chen, Flum and Grohe \cite{TCS:CFG05} 
and Buss and Islam \cite{TCS:BI06} to
do this (details of this model and its relationship to parameterized
complexity are given in Appendix A).
To transform the algorithm into an aftp-program showing that the 
decision problem
is in the class $W[P]$, we simply replace the enumeration by an
existential steps guessing the sets of indices of size $k$ giving
the support of the strategies of Player 2 and Player 3.
In the remainder of this section we will show that for the special
case 0-1 payoffs the decision problem is in the class $W[1]$. The idea
is to precompute, for every possible $k \times k \times k$ game with
0-1 payoffs, whether the minmax value for Player 1 is at most
$\alpha$. As in the $W[P]$ case, indices of support of
strategies is computed but now the $k \times k \times k$ subgame is
used as an index in the precomputed table. To see that this can be
turned into an appropriate aftp-program, we will formally define the
relations used.

Assume that the payoffs of Player 1 are given as a $k$-tuple of $m
\times n$ 0-1 matrices $(U^1,\dots,U^k)$. Define a unary relation $A$
over $k$-tuples of $k \times k$ 0-1 matrices as follows.
$(M^1,\dots,M^k) \in A$ if and only if the minmax value for Player 1
in the $k \times k \times k$ subgame given by $(M^1,\dots,M^k)$ is at
most $\alpha$.

Define a $4$-ary relation $B$ having as first argument a $k$-tuple of
$k \times k$ 0-1 matrices and with the last 3 arguments being indices from $1$
to $k$ as follows.
\[
((M^1,\dots,M^k),l,i,j) \in B \quad \text{if and only if} \quad M^l_{ij} = U^l_{ij} \enspace .
\]
The algorithm first computes the relations $A$ and $B$. In the
guessing steps the algorithm guesses a $k$-tuple of matrices
$(M^1,\dots,M^k)$ and indices $i_1,\dots,i_k$ and $j_1,\dots,j_k$. The
final checks the algorithm must perform are
$(M^1,\dots,M^k) \in A$ and 
$((M^1,\dots,M^k),l,i_a,j_b) \in B$
for all $l,a,b \in \{1,\dots,k\}$. The number of steps used for the
guessing the indices and the final check is a function depending only
on the parameter $k$ as required.

As discussed by Buss and Islam \cite{TCS:BI06} we can in a generic way
transform an algorithm utilizing a constant number of relations such
that it only utilizes a single binary relation, thereby obtaining
an aftp-algorithm showing that the decision problem is in $W[1]$.
\subsection{Proof of Theorem \ref{thm:reduction}}



Before starting the proof, we remark that the reduction is based
on similar ideas as the reduction proving {\bf NP}-hardness of the
problem by
Borgs {\em et al.} However, they reduce from 3-coloring rather
than clique, and in their coloring based games, we don't see how to
restrict the strategy space of Player 1 to a small number of
strategies, so as to obtain fixed-parameter intractability.

We now describe the reduction.
Given an undirected graph $G = (V,E)$, with $|V| = n$. We construct a
$2 k \times k n \times k n$ game from $G$ and compute the threat
value to within $1/(4k^2)$. We construct the game in the following
way. Let $A_1 = \{1,\dots,k\}\times \{2,3\}$ be the strategy space of
Player $1$ and $A_2 = A_3 = \{1,\dots,k\} \times V$ be the strategy
space of Player $2$ and Player $3$. We define the payoff of Player $1$ as:
\[
u_1((x_1,i),(x_2,v_2),(x_3,v_3)) = \begin{cases}
1 & \text{if $x_1 = x_i$}\\
1 & \text{if $x_2 = x_3$ and $v_2 \neq v_3$}\\
1 & \text{if $x_2 \neq x_3$ and $v_2 = v_3$}\\
1 & \text{if $v_2 \neq v_3$ and $(v_2,v_3) \notin E$}\\
0 & \text{otherwise}
\end{cases}
\]
Player $2$ and Player $3$ will try to minimize the payoff of Player $1$,
hence we shall call them {\em bullies}.
One can think of the game as the bullies each choosing a
label and a vertex of $G$. Player $1$ then chooses one of the bullies
and tries to guess his label. If he guesses correctly he will get a
payoff of $1$. If not, he will get a payoff of $0$, unless the 
bullies do one of the following:
\begin{itemize}
\item[(i)]
  Choose the same label, but different vertices.
\item[(ii)]
  Choose different labels, but the same vertex.
\item[(iii)]
  Choose a pair of distinct vertices that does not correspond to an
  edge.
\end{itemize}
In which case he will get a payoff of $1$. The intuition behind the
proof is that the bullies will be able to avoid these three cases
if the graph contains a $k$-clique, thereby better punishing Player
$1$. 

We first notice that if $G$ contains a k-clique, the bullies can bring
down the payoff of Player $1$ to $1/k$ by choosing a vertex
from the k-clique uniformly at random and agreeing on a labeling of
the vertices. 
Let $p_{\max}$ be the highest probability that any of the bullies 
choose any label $j\in\{1,\dots,k\}$. Player $1$ will, on the other
hand, always be able to get a payoff of $p_{\max}$ by choosing $j$ and
the corresponding player. It follows that the threat value of the game
is $1/k$ when $G$ contains a k-clique.

Next, we will show that if $G$ contains no k-clique, the bullies will
at most be able to bring down the payoff of Player $1$ to $1/k +
1/(4k^2)$. We will do a proof by contradiction, so lets assume that the
bullies can force Player $1$ to get a payoff less than $1/k +
1/(4k^2)$ and show that this can never be the case. We have already seen
that in this case $p_{\max} < 1/k + 1/(4k^2)$.

Consider the case where Player $1$ always chooses Player $2$ and
guesses a label uniformly at random. In this case Player $1$ will
always guess the correct label with probability $1/k$ independently of
the actions of the bullies. Hence, we need to show that for $p$ being the
probability of (i), (ii) or (iii)
happening we have that:

\[
\frac{1}{k} + \left(1 - \frac{1}{k}\right) p <
\frac{1}{k} + \frac{1}{4k^2} \Rightarrow
p < \frac{1}{4k(k-1)}
\]

In
particular the probability of (i) happening is less than $1/(4k(k-1))$. Let
$p_{\min} \geq 1 - (k-1)p_{\max} > 3/(4k) + 1/(4k^2)$ be the minimum
probability assigned to any label by either of the bullies. Let
$(x,v)_i$ denote that Player $i$ chooses the label $x$ and the vertex
$v$. We will use $\cdot$ as a wildcard, such that $(x,\cdot)_i$ means
that Player $i$ chooses the label $x$ and $(\cdot,v)_i$ means
that Player $i$ chooses the vertex $v$. For $v,w\in V$ we see that:
\begin{align*}
\frac{1}{4k(k-1)} &> \Pr\bigl[x_2 = x_3 \text{ and } v_2 \neq v_3\bigr] \\
&= \sum_{j=1}^k \sum_{v \neq w} 
\Pr\bigl[(\cdot,v)_2\mid(j,\cdot)_2\bigr]\Pr\bigl[(j,\cdot)_2\bigr]\Pr\bigl[(\cdot,w)_3\mid(j,\cdot)_3\bigr]\Pr\bigl[(j,\cdot)_3\bigr]
\\
&\geq p_{\min}^2 \sum_{j=1}^k \sum_{v \neq w} 
\Pr\bigl[(\cdot,v)_2\mid(j,\cdot)_2\bigr]\Pr\bigl[(\cdot,w)_3\mid(j,\cdot)_3\bigr]
\\
&=
p_{\min}^2 \sum_{j=1}^k \left( 1 - \sum_{v \in V} 
\Pr\bigl[(\cdot,v)_2\mid(j,\cdot)_2\bigr]\Pr\bigl[(\cdot,v)_3\mid(j,\cdot)_3\bigr]\right)
\\
&= k p_{\min}^2 - p_{\min}^2 \sum_{j=1}^k \sum_{v \in V}
\Pr\bigl[(\cdot,v)_2\mid(j,\cdot)_2\bigr]
\Pr\bigl[(\cdot,v)_3\mid(j,\cdot)_3\bigr].
\end{align*}
This implies that $\forall l \in \{1,\dots,k\}:$
\begin{align*}
& \sum_{v \in V}
\Pr\bigl[(\cdot,v)_2\mid(l,\cdot)_2\bigr]
\Pr\bigl[(\cdot,v)_3\mid(l,\cdot)_3\bigr]\\
 &>
k - \frac{1}{4k(k-1) p_{\min}^2} - \sum_{j \neq l} \sum_{v \in V}
\Pr\bigl[(\cdot,v)_2\mid(j,\cdot)_2\bigr]
\Pr\bigl[(\cdot,v)_3\mid(j,\cdot)_3\bigr]\\
&\geq 1 - \frac{1}{4k(k-1)p_{\min}^2}
>1 - \frac{1}{4k(k-1)\left(\frac{3}{4k} + \frac{1}{4k^2}\right)^2}\\
&= 1 - \frac{4k^3}{(k-1)(3k+1)^2} > \frac{1}{2}, \text{ for } k \geq 5
\end{align*}

Let $v_i^j$ be the vertex chosen with highest probability by player
$i$ given that he chooses label $j$. The above implies that for $k
\geq 5$ we have for all
$j \in \{1,\dots,k\}$ that $v_2^j = v_3^j$, and we will
therefore simply refer to this vertex as $v^j$. We also
see that for $k
\geq 5$:

\begin{align*}
\forall i \in \{2,3\}, \forall j \in \{1,\dots,k\}: \Pr\bigl[(\cdot,v^j)_i\mid(j,\cdot)_i\bigr] > \frac{1}{2}
\end{align*}

That is, for every label $j$ the bullies agree on some vertex $v^j$ that they choose
with high probability when choosing $j$. We will use this to get a
contradiction from case (ii) or (iii). For $j,l \in \{1,\dots,k\}$, it
will either be the case that there exists some $v^j = v^l$, with $j
\neq l$, or that all the $v^j$'s are distinct. In the first case
Player $1$ will, with high probability, get a payoff of $1$ when one
of the bullies chooses label $j$ and the other chooses label $l$
(case (ii)). In the second case there will exist a
pair of distinct labels $j$ and $l$, such that there is no edge
between $v^j$ and $v^l$, since the graph doesn't contain a
k-clique. Hence, this will cause Player $1$ to get a payoff of $1$,
with high probability, when one
of the bullies chooses label $j$ and the other chooses label $l$ (case
(iii)). In both cases we get that for $k \geq 5$:

\[
\sum_{i=0}^1
\Pr\bigl[(\cdot,v^j)_{2+i}\mid(j,\cdot)_{2+i}\bigr]\Pr\bigl[(j,\cdot)_{2+i} \bigr]
\Pr\bigl[(\cdot,v^l)_{3-i}\mid(l,\cdot)_{3-i}\bigr]\Pr\bigl[(l,\cdot)_{3-i} \bigr]
\]
\[
> 2 p_{\min}^2 \left( \frac{1}{2} \right)^2
> \frac{1}{2} \left(\frac{3}{4k} + \frac{1}{4k^2}\right)^2 
= \frac{(3k+1)^2}{32 k^4}
> \frac{1}{4k(k-1)}
\]

Which contradicts that this should happen with probability less than
$1/(4k(k-1))$, meaning that if $G$ contains no k-clique the bullies will
at most be able to bring down the payoff of Player $1$ to $1/k +
1/(4k^2)$, which completes the reduction.

%
%
%

\section{Conclusions and open problems}
As mentioned above, an important open problem is achieving a
non-trivial approximation of the minmax value of an 
$n \times n \times n$ game in polynomial, rather than quasi-polynomial
time. Another interesting question comes from the following notions: 
The {\em threat point} of a game is defined to be the vector of
minmax values for each of its three players. We may consider
approximating the threat point of a game where one of
the players has few strategies. For this, we have to consider 
the problem of approximating the threat value for
Player 1 in a three-player $n \times k \times n$ game. That is, 
it is now one of the two ``bullies'' rather than
the threatened player that has few
strategies. We observe that for constant $\epsilon > 0$ such an
approximation
can be done efficiently by
simply discretizing the mixed strategy space of the player with
few strategies using a lattice with all simplex points having distance
at most $\epsilon$ to some lattice point, and then 
for each lattice point solving the game for the
remaining two players using linear programming.
Combining this with Theorem \ref{thm:support} gives us the following corollary
\begin{corollary}
There is an algorithm that, given an $n \times k \times n$ game 
with 0-1 payoffs and an $\epsilon > 0$, computes
the threat point within additive error $\epsilon$ in time
$(n/\epsilon)^{O(k)}$.
\end{corollary}
The discretization technique gives algorithms with poor dependence
on the desired additive approximation $\epsilon$. 
We leave as an open problem if
the minmax value of an $n \times k \times n$ 0-1 game can be approximated
within $\epsilon$ in time $(n \log(1/\epsilon))^{O(k)}$.
\subsection*{Acknowledgements}
We would like to thank Bernhard von Stengel for valuable help.
\bibliographystyle{abbrv}

\pagebreak
\section*{Appendix A: Parameterized complexity}
For the convenience of the reader, we summarize notions from
the theory of parameterized complexity used in this paper.
\begin{definition}
  A parameterized problem is a set of pairs $Q \subseteq \Sigma^*
  \times \Sigma^*$, where $\Sigma$ is a finite alphabet. The second
  coordinate is the \emph{parameter}. 
\end{definition}
We may think of $Q$ as being a subset of $\Sigma^* \times \N$, and we
will do so henceforth.
\begin{definition}
  A parameterized problem $Q \subseteq \Sigma^* \times \N$ is
  \emph{fixed parameter tractable}, if there is a computable function
  $f : \N \rightarrow \N$, a polynomial $p$ and an algorithm that
  given $(x,k) \subseteq \Sigma^* \times \N$ decides if $(x,k)
  \in Q$ in at most $f(k)p(|x|)$ steps.
\end{definition}
Let $\FPT$ be the class of parameterized problems that are fixed
parameter tractable. To compare the complexity of problems that are
not fixed parameter tractable the notion of fixed parameter reductions
are introduced.
\begin{definition}
  An fpt-reduction from a parameterized problem $Q \subseteq \Sigma^*
  \times \N$ to a parameterized problem $Q' \subseteq (\Sigma')^*
  \times \N$ is a function $R : \Sigma^* \times \N \rightarrow
  (\Sigma')^* \times \N$ such that
\begin{enumerate}
\item For all $(x,k) \in \Sigma^* \times \N$ we have $(x,k) \in Q$ if
  and only if $R(x,k) \in Q'$.
\item There exists a computable function $g : \N \rightarrow \N$ such
  that for all $(x,k) \in \Sigma^* \times \N$ we have $k' \leq g(k)$,
  where $R(x,k)=(x',k')$.
\item There exists a computable function $f : \N \rightarrow \N$ and a
  polynomial $p$ such that $R$ can be computed in at most $f(k)p(|x|)$
  steps.
\end{enumerate}
\end{definition}

A hierarchy of parameterized complexity classes (the $W$ hierarchy)
\[
\FPT \subseteq W[1] \subseteq W[2] \subseteq \dots \subseteq \WSAT
\subseteq W[P]
\]
was defined by Downey and Fellows \cite{SICOMP:DF95}. The classes
$W[t]$ were defined by the class of problems fpt-reducible to the
weighted satisfiability problem for constant depth, so-called
\emph{weft}-t circuits. The class $\WSAT$ is the class of problems
reducible to the weighted satisfiability problem for Boolean formulas,
and finally $W[P]$ is the class of problems reducible to the weighted
satisfiability problem for Boolean circuits.

Chen, Flum and Grohe \cite{TCS:CFG05} gave a machine based
characterization of all classes in the $W$ hierarchy, except $\WSAT$,
that we will use instead. The remainder of this section is a
description and statement of this characterization.

A WRAM is an alternating random access machine. Its registers are
partitioned into two sets: \emph{standard registers} $r_0,r_1,\dots$,
and \emph{guess registers} $g_0,g_1,\dots$\enspace. The guess
registers will be indirectly indexed by the standard registers. We
write $g(r_i)$ for $g_{r_i}$. For accessing the standard registers the
machine has all the instructions of a standard deterministic random
access machine. For accessing the guess registers the machine has the
following instructions.

\begin{center}
\begin{tabular}{l|l}
  \hline
  Instruction & Semantics \\
  \hline
  EXISTS $\uparrow j$ & \emph{existentially} guess a natural number $\leq r_0$ and store it in $g(r_j)$.\\

  FORALL $\uparrow j$ & \emph{universally} guess a natural number $\leq r_0$ and store it in $g(r_j)$.\\
  JGEQUAL $i\ j\ c$   & if $g(r_i) = g(r_j)$ jump to the instruction with label $c$.\\
  JGZERO $i\ j\ c$    & if $r_{\left<g(r_i),g(r_j)\right>} = 0$ jump to the instruction with label $c$.\\
  \hline
\end{tabular}
\end{center}
Here $\left< \cdot , \cdot \right>$ is any simple pairing function of
natural numbers with $\left<i,j\right>\leq (1+\max(i,j))^2$ and
$\left<0,0\right>=0$.

A program $P$ for a WRAM is an afpt-program, if there exists a
computable function $f$ and a polynomial $p$ such that for every input
$(x,k)$ and every possible run, the program
\begin{itemize}
\item performs at most $f(k)p(|x|)$ steps.
\item performs at most $f(k)$ existential and universal steps.
\item accesses only the first $f(k)p(|x|)$ standard registers.
\item only stores numbers $\leq f(k)p(|x|)$ in any register.
\end{itemize}
With these definitions we have the following characterization of the
classes $W[t]$ and $W[P]$.
\begin{theorem}
  A parameterized problem $Q$ is in $W[P]$ if and only if there is an
  aftp-program without universal instructions deciding $Q$. This also
  holds if we allow the aftp-program direct access to the guess
  registers.
\end{theorem}
\begin{theorem}
  Let $Q$ be a parameterized problem and $t \geq 1$. Then $Q$ is in
  $W[t]$ if and only if there is a computable function $h$, a natural
  number $u$ and an aftp-program deciding $Q$ such that every run of
  the program on any given input $(x,k)$ satisfies.
\begin{itemize}
\item All nondeterministic steps are among the last $h(k)$ steps of the computation.
\item The first nondeterministic step is existential.
\item There are at most $t-1$ alternations between existential and
  universal states.
\item Any sequence of steps without alternations, except the first
  existential sequence, contains at most $u$ nondeterministic steps.
\end{itemize}
\end{theorem}
\end{document}